\documentclass[journal=jacsat,manuscript=article]{achemso}
\usepackage[T1]{fontenc} 
\usepackage{epsfig}
\usepackage{subfig}

\author{Ra\'ul Fuentes-Azcatl}
\altaffiliation{Departamento SIC. Universidad Aut\'onoma Metropolitana, Lerma. Av. de las Garzas 10, El pante\'on, 52005 Lerma de Villada, Edomex. M\'exico.}
\email{r.fuentes@correo.ler.uam.mx}

\author{Jos\'e Rafael Bordin}
\affiliation{Departamento de F\'isica, Instituto de F\'isica e Matem\'atica, Universidade Federal de Pelotas. Caixa Postal 354, 96001-970, Pelotas, Brazil.}
\email{jrbordin@ufpel.edu.br}

\author{Marcia C. Barbosa}
\email{marcia.barbosa@ufrgs.br}
\affiliation{Instituto de F\'isica, Universidade Federal do Rio Grande do Sul, Caixa Postal 15051, CEP 91501-970, Porto Alegre, RS, Brazil}

\title {Water inside charged nanoslits: Structure and Dielectric  study with a novel water model FAB/$\epsilon$ }

\begin{document}

\begin{abstract}
	In this work, the dielectric behavior of water inside charged nanoslit of graphene is studied to analized the water molecules under electrical confinement; through polarizing the nanoslit of graphene, creating an electric field inside the nanopore. How the water molecules are structured under this type of electrical confinement is studied with two force fields of water, the three-site water models here used are the SCP/$\epsilon$ and the FAB/$\epsilon$, the first is a rigid model that improves the SPC model and the second is a flexible model that improves all the force fields of three sites non-polarizables and flexibles.
	
\end{abstract}

\section{Introduction}
Water is unique not only because life, as we known, happens in water but also due to its relevance in industry. It exhibits more than 70 known anomalous behavior~\cite{url}, and the water behavior becomes even more stranger when confined in pores with few nanometers \cite{Zangi2004, Gao18, Santi21}. Specially, the comprehension of confined water is important from many aspects. Water  confined inside biological proteins~\cite{Lynch20}, in the DNA grooves~\cite{passos} or inside rocks nanopores, as it is in the shale formation process~\cite{Sun18} presents additional anomalies and modify in a non trivial way the confined media. Also, many new technologies have been proposed based in the water behavior at this scale~\cite{Bampo18}. Among the many nanomaterials employed in these new technologies~\cite{Shen21}, single layers of graphene (SLG) emerge as one of the most prominent material~\cite{Naushad}. It was successfully used in a wide range of applications such as water filtration \cite{ wei18, Xu:2017aa}, single-molecule sensors  \cite{Geng, Valenga20},  ion  selectivity  \cite{Thomas, Bordin,Andreeva21}, and  energy  conversion  and storage \cite{Park, Liu19, Olabi21,graf-co2e}. 

Water ordering and orientation at the nanopore interface play a major role in the nanofluidic research and all the possible applications. For instance, recent experimental~\cite{Suga19, Fumagalli18} and computational~\cite{Zhang13,Zhu21, Mote20, Mote20b,Olivieri21} studies have indicated an confinement induced reduction in the water dielectric constant that can be related to the water structure at the interface or to long-ranged anisotropic dipole correlations combined with an excluded-volume effect of the low-dielectric pore material. Also, the surface layer structure can drastically affect the dynamical behavior of nanoconfined water~\cite{C7CP02058A,Celebi19}, in fact, the dynamics and the static dielectric constant are related as well~\cite{Mondal20}. Therefore, understand the water behavior at the solid boundaries under extreme confinement is of great importance (and a great challenge) to fundamental research~\cite{Cerveny16} and technological applications.

Despite the huge advances in nanotechnology, experiments of fluids confined at the length scale of few nanometers remain as a considerable challenge~\cite{Sam20,xu20, Hong21}. Is this sense, the use of Molecular Dynamics (MD) simulations arises as a suited approach to study nanofluidic systems~\cite{gravelle:hal-02375018}. Simulations are used to interpret  experiments, but also to study properties that can not be accessed directly by them. The proper depiction by MD simulations relies on the definition of the Classical Force Fields based in additive pairwise potentials for the particle particles interaction~\cite{Allen89,Underwood2018}. 

In the specific case of water, a large number of atomistic models have been proposed. Most of them are rigid nonpolarizable models, with 3 or more sites.~\cite{tip4pe,spce,tip4peMarcia} Even this simple models can reproduce most of the bulk water behavior, there are recent models that can reproduce the structure and infrared spectrum as the FBA/$\epsilon$ and TIP4P/$\epsilon$ flexible~\cite{fbae,tip4pef}, improving all current non-polarizable and flexible models~\cite{tip4pef}. Properly include polarization in classical simulations remain as an huge task, but efforts are being done in this direction in the recent years~\cite{doi:10.1021/ct900576a, HALGREN2001236, doi:10.1021/ja00007a021}, but the flexibility offers a type of polarization to the molecule giving freedom of movement to the charge, as in the model FBA/$\epsilon$~\cite{fbae}.

Also, the addition of the molecule flexibillity is important particularly in the study of water solutions. For instance, the H--O--H angle is 104,474$^o$ for water molecules in the gas phase, with a distance of 0.095718 nm between oxygen and each hydrogen. However, in the liquid phase at 298 K and 1 bar, the average angle changes to 106$^o$~\cite{waterbook}. Once each water molecule, in bulk, form up to four hydrogen bonds, small variations in the H--O--H angle can be relevant to the molecules cluster structure. Under extreme confinement, where the hydrogen bond is suppressed by the geometrical constraints, variations in the angle can be even more relevant for the confined water structural, dynamic, thermodynamic and dielectric properties. 

A drawback of the classical atomistic models fitted using bulk properties is that distinct classical models, even though agree in some measures in the confined geometries, lead to different quantitative results in other quantities~\cite{Tsim19, Tsim20, Abal20}. For instance, Celebi and co-workers~\cite{Celebi19} recent analyzed the slip length of water inside graphene nanopores employing rigid non-polarizable models and found considerable differences and deviations in the viscosity and slip length when compared with experimental results. With this, a question that naturally arises is how the molecule flexibillity can improve the reproduction of the behavior of nanoconfined water when compared with the classical atomistic approach.

In order to answer this question we here study the behavior of two water models confinement between two parallel single layers graphene (SLG) sheets. The SLG walls can be neutral, modeling a nanopore made by pristine graphene, or charged, as a minimal model for a nanocapacitor. This charged system is not only interesting from the practical applications of nanocapacitors, but to understand how the charged wall and electric field, modify the structure of the water and how it will reflect in the fluid behavior.\\
  We compare the behavior under confinement of  two recent three site water models that improved the traditional SPC water model~\cite{grigera87}. The rigid SPC/$\epsilon$ model~\cite{spce}, was constructed based on the SPC model, but show a with better depiction of the dielectric behavior. More recently, the flexible FAB/$\epsilon$ \cite{fbae} model was constructuded to enhance the thermodynamic properties at different pressure and temperature  conditions in comparison to all three-site models by adding extra degrees of fredom to the sites. With this, we can analyze how variations in the internal angle and bond lengths will affects the behavior of the interfacial and bulk-like water inside neutral and charged nanopores. 

Our paper is organized as follows. In Section 2 we introduce the water and nanopore models and the details about the simulation method. In Section 3 we show and discuss our results, and the conclusions are presented in Section 4.

\section{The Models}
\label{model}
\subsection{The Water Models}

In this work we compare two water models, one rigid and one fully flexible. The rigid model, namely SPC/$\epsilon$ \cite{spce} model, is an improved SPC model with better depiction of the dielectric behavior which helped to have better force fields of monovalent ions \cite{nacle,kbre}. The another water model is the recently proposed FAB/$\epsilon$ \cite{fbae}, flexibility helps to study changes in structure due to other substances such as ionic liquids \cite{IL-h2o}, fully flexibility of the model: it has harmonic potentials at the O-H$_{bond}$ and at the angle formed by the three molecule site,which helps to improved the force field as in the CO$_2$ molecule \cite{co2e}.\\

 While in the SPC/$\epsilon$ model the atoms in each molecule are maintained connected by rigid constraints, in the FAB/$\epsilon$ flexible model there is the addition of the harmonic potential in the O-H bond,
\begin{equation}
\label{k}
U_k(r)=\frac{k_r}{2}(r-r_0 )^2 
\end{equation}
\noindent and in the H-O-H angle,
\begin{equation}
\label{theta}
U_{\theta}(\theta)=\frac{k_{\theta}}{2}(\theta-\theta_0)^2 ,
\end{equation}
\noindent where $r$ is the bond distance and $\theta$ is the bond angle. The subscript $0$ denotes their equilibrium  values, and $k_r$ and $k_{\theta}$ are the corresponding spring constants. 

For the intermolecular potential between two molecules the LJ and Coulomb interactions are used for non-polarizable models,

\begin{equation}
\label{ff}
u(r) = 4\epsilon_{\alpha \beta} 
\left[\left(\frac {\sigma_{\alpha \beta}}{r}\right)^{12}-
\left (\frac{\sigma_{\alpha \beta}}{r}\right)^6\right] +
\frac{1}{4\pi\epsilon_0}\frac{q_{\alpha} q_{\beta}}{r}
\end{equation}

\noindent where $r$ is the distance between sites $\alpha$ and $\beta$, $q_\alpha$ is the electric charge of site $\alpha$, $\epsilon_0$ is the permitivity of vacuum,  $\epsilon_{\alpha \beta}$ is the LJ energy scale and  $\sigma_{\alpha \beta}$ the repulsive diameter for an $\alpha-\beta$ pair. The cross interactions between unlike atoms are obtained using the Lorentz-Berthelot mixing rules,

\begin{equation}
\label{lb}
\sigma_{\alpha\beta} = \left(\frac{\sigma_{\alpha\alpha} +
	\sigma_{\beta\beta} }{2}\right);\hspace{1.0cm} \epsilon_{\alpha\beta} =
\left(\epsilon_{\alpha\alpha} \epsilon_{\beta\beta}\right)^{1/2}
\end{equation}

The table~\ref{table2} shows the force field values of the models mentioned above.

\begin{table}
	\caption{Parameters of the three-site water models considered in this work.
	}
	\label{table2}
	\begin{tabular}{|ccccccccc|}
		\hline\hline
		model	&	$k_{b}$	&	$r _{OH}$ 	&	$k_{a}$	&	$\Theta$ 	&	$\varepsilon_{OO}$	&	$\sigma_{OO}$	&	$q_{O}$	&	$q_{H}$	\\

		&	kJ/ $mol$ {\AA}$^{2}$ 	&	{\AA}	&	kJ/ $mol$ rad$^{2}$	&	deg	&	kJ/mol	
		&	{\AA}	&	e	&	e	\\

		SPC/$\epsilon$	\cite{spce}&	-	&	1.000	&	-	&	109.45	&	0.705859	&	3.1785	&	-0.8900	&	0.4450	\\

		FBA/$\epsilon $	\cite{fbae}&	3000	&	1.027	&	383	&	114.70	&	0.792324	&	3.1776	&	-0.8450	&	0.4225	\\
		
		\hline
	\end{tabular}
\end{table}

\subsection{The Graphene Model}
\noindent 

The graphene walls were modeled as a highly oriented pyrolytic graphite (HOPG) sheet. It is a highly pure and ordered form of synthetic graphite. It is characterized by a low mosaic spread angle, meaning that the individual graphite crystallites are well aligned with each other. The original force field for the carbons in this graphene model is composed only by a Lennard-Jones interaction \cite{Bonthuis11}. Here, charges are included in every carbon atom to have an uniform surface charge densities, $\sigma_S$, as given by Shim et al. \cite{Shim11}. The charges are increased from zero (neutral confinement) to create the desired voltage between the walls.                  

\begin{table}[h]
	\caption{Force field parameters of grafene model.}
	\label{grafene}
	\begin{center}
		\begin{tabular}{|cccc|}
			\hline
			\hline Model  & $q_C/e$& $\sigma$/\AA & $(\epsilon/k_B)$/K\\
			\hline
			
			C  & 0.0143$^\dagger$ &  3.401 & 35.46\\
			
			\hline
		\end{tabular}\\
		$\dagger$-example of charge for a uniform surface charge densities of $\sigma_S$=  0.86e/nm$^2$
		\end {center}
	\end{table}
	
	\section{Simulation Details}
	
	\noindent We employed Molecular Dynamics (MD) simulation in the $NVT$ ensemble to study the behavior of the confined water inside a nano-capacitor. The nano-capacitor was constructed with two parallel SLG with 240 carbon atoms placed in an area L$_x$ =2.39315 L$_y$ = 2.39315 nm separated by a distance of L$_z$ = 2.85 nm. Then, 458 H$_2$O molecules were initially placed randomly inside the nano-capacitor space. Simulations with  electric potential ranging from 0 to 150 V were performed. The electrical potential was created by inducing a surface charge density in the SLG parallel plates by including point charges in each carbon atom in order to get the desired value. The left wall is the positive plate and the right wall the negative one. The net charge of the system is zero.
	
	\noindent All simulations were conducted using the molecular dynamics package GROMACS 2016\cite{Pall15}. Periodic boundary conditions were imposed in the $xy$-plane. The equations of motion were solved using the leap-frog algorithm~\cite{Allen89,Pall15} with a time step of 1 fs. The temperature was set to $T =$ 298 K with the Nos\'{e}-Hoover thermostat~\cite{Tuckerman01} with coupling parameter 0.6 ps. The LINCS algorithm was employed for the ridig bons. The Lennard-Jones  potential was truncated at 10 \AA without any long-range correction in energy or pressure. The real part of the Coulomb potential was also truncated at 10 \AA. The Fourier component of the Ewald sums was evaluated with the particle mesh Ewald (PME) method \cite{Essman95} using a grid spacing of 0.35 \AA \ and a three degree polynomial for the interpolation. All simulations were run for 50 ns, were the first 5 ns are the equilibration time and the next 45 ns the production time, with a output frequency for positions and velocities at each 1000 steps.

	\section{Results}
	
	\setcounter{subfigure}{0}
	 
	\begin{figure}[]
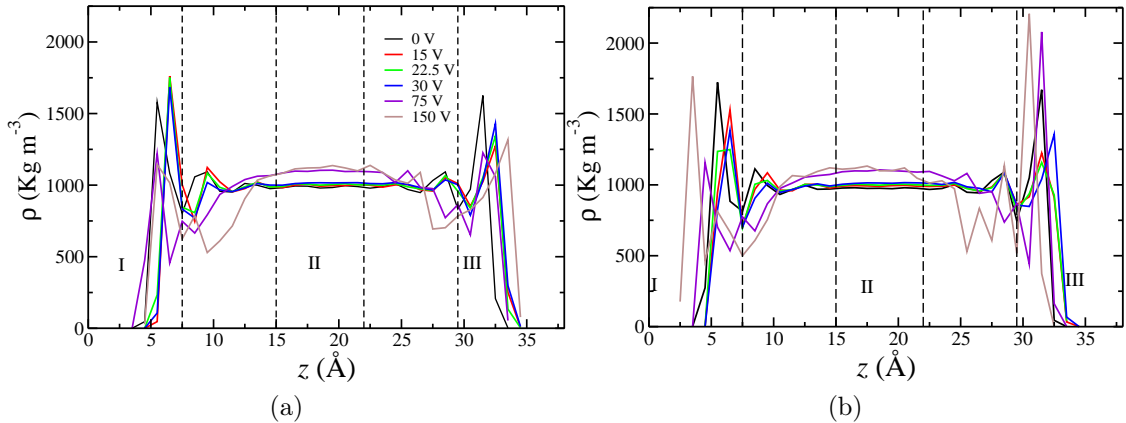

		\subfloat[]{\includegraphics[width=0.45\textwidth]{fab-dens.eps}}		\subfloat[]{\includegraphics[width=0.45\textwidth]{spce-dens.eps}}
\caption{Density profile along the Z direction for the FAB/$\varepsilon$ model (a) and for the SPC/$\varepsilon$ model (b) for various potentials between the plates. Three regions are defines: region I, near the left wall, region II, the bulk-like region at the center, and region III, near the right wall.}
		\label{Fig1}
	\end{figure}
	
	\noindent Water (and fluids in general) under confinement assumes a layered structure near the walls and have a bulk-like region in the nanopore center~\cite{Sobrino16,Zokaie15, Gao18, KoB13, KoB14, KoB15}. We observe the same behavior for both water models as illustrated by the density profiles along the $z$-direction for the distinct applied voltages are shown in figure \ref{Fig1}(a) for the flexible FAB/$\varepsilon$ model and in graph \ref{Fig1}(b) for the rigid the SPC/$\varepsilon$ model. The water central density is the value expected by the  bulk H$_2$O density of each model for the specific temperature and pressure, and the density profiles are similar to the observed in previous works for water confined by neutral and charged nanoplates \cite{Gao18, berkowitz,Harrach14, Celebi18, Zhao20}. The same trends are observed for both H$_2$O models. To understand the distinct water behaviors inside the nanopore, we divided it in three regions: the contact layers in the region I - near the positive wall - and in the region III - at the negative wall and the bulk-like region II.
	
	
	\begin{figure}[]
		\centering
		\subfloat[]{\includegraphics[width=0.45\textwidth]{mol-V.eps}}
		\qquad
		\subfloat[]{\includegraphics[width=0.45\textwidth]{mu-V.eps}}
		
		\subfloat[]{\includegraphics[width=0.45\textwidth]{Ang-V.eps}} 
		\qquad
		\subfloat[]{\includegraphics[width=0.45\textwidth]{bondOH-V.eps}}
		\qquad
		\subfloat[]{\includegraphics[width=0.4\textwidth]{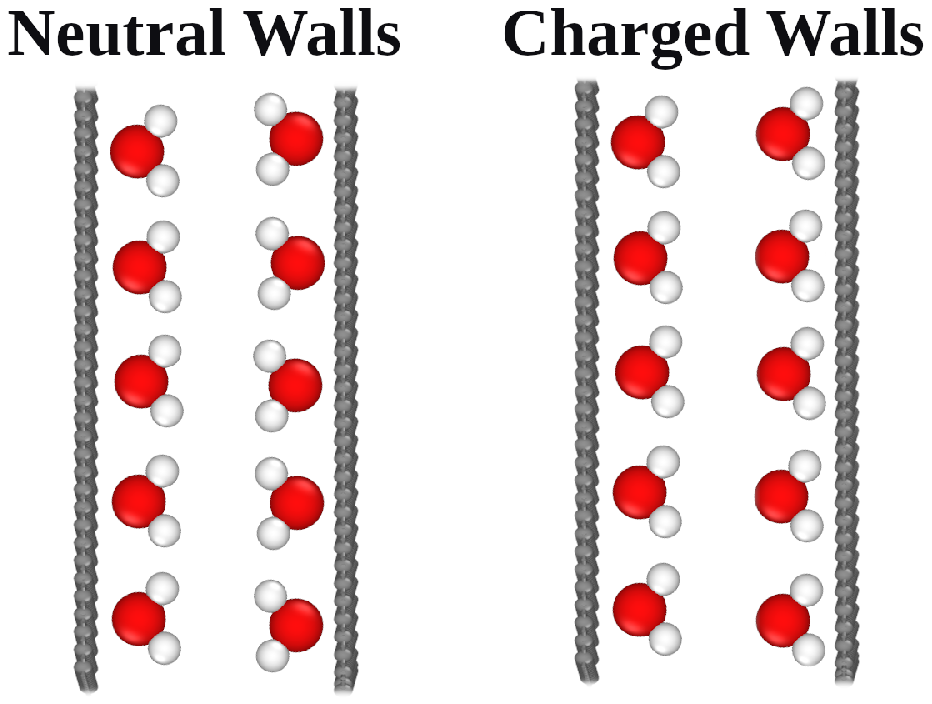}}
		\caption{(a) Dipole moment (b)H--O--H average angle  (c) O-H average length for confined H$_2$O  versus voltage in the distinct regions between the plates for the SPC/$\varepsilon$ rigid  model (square)
			) and FAB/$\varepsilon$ flexible (circles) water models. (d) Schematic view of the water near the neutral nanopore walls and close to the charged walls.}
		\label{Fig2}
	\end{figure}

	The distinct occupancy and layering in the regions defined by the slabs I, II and III can lead to distinct physical properties for the water molecules. In order to understand the role of  model flexibility in water properties in each region we evaluated the molecule dipole moment $\mu$ and variation in the internal bonds and angles. These quantities are constant for rigid models and varies for flexible models. The dipole moment versus voltage for the different regions I, II and III, are shown in figure \ref{Fig2}(a), while the H--O--H internal angles and the H--O bond length are shown in the figures \ref{Fig2}(b) and (c), respectively. The results indicates that for the FAB/$\varepsilon$ model $\mu$ changes as the voltage increases in a non monotonic way. In the region II, $\mu$ presents values slightly above the bulk with a small increase with the voltage.  In the region I, $\mu$ increases with the voltage. In the region II, $\mu$ decreases with the voltage below a threshold, $V<V^*$ ($V^*$ is a threshold voltage),  and increases above it. 
	
	The behavior of $\mu$ correlates with changes in the H-O-H angle, $\theta$ and the O-H$_{bond}$ length as shown in figure~\ref{Fig2}(b) and (c). Near the cathode, in the slab region I, $\theta$  increases as $V$ increases for $V<V^*$, a consequence of the repulsion between the H's atoms and the positive wall. Likewise, this repulsion compress the hydrogens in direction to the O atom, leading to a decrease of the O-H$_{bond}$ length for $V<V^*$, as shown in the figure \ref{Fig2}(c). These structural changes lead to the decrease in $\mu$ observed in the region I,  $V<V^*$  in figure  \ref{Fig2}(a). Above this voltage threshold, $\mu$ increases with $V$. This change in the slope is followed by the structural changes. As $V$ grows above this threshold, $\theta$ decreases and the O-H$_{bond}$ length increases. 
	
	The inflection point in $V=V^*$ coincides with the voltage where the peak in the density profile moves closer to the charged wall, as we can see in figure \ref{Fig1}(a) for $V = 75$ V and 150 V. This indicates that the negatively charged O atoms are now closer to the wall, increasing the wall repulsion to the H atoms and increasing the bond length and the bond angle.

	Near the anode, in the region III, we do not see the monotonic behavior of $\mu$, $\theta$ and the H--O bond length. $\mu$ increases with $V$, as the bond length also increases and $\theta$ decreases, reinforcing the relations observed near the positive plate. Here, also the increase in $\mu$ is consequence of the molecules going closer to the negative wall. However, here the H atoms are attracted. As the graphene wall pull them in her direction, the bond length increases and $\theta$ decrease. The elongation in the bond length lead to the increase in the dipole moment. Also, is interesting to mention that the confinement changes the bond length (and $\mu$) even for the bulk-like region II and neutral walls case. As reported by Fuentes and Barbosa~\cite{fbae}, the bond length for the FAB/$\varepsilon$ in the liquid phase at 298 K and 1 bar is 0.9495 \AA. However, for zero volts, the average bond length in the region II is 1.058 \AA, this indicates that the molecule internal structure in the bulk-like regime is affected by the confinement.

	\begin{figure}[t]
		\vspace{2.0 cm}
		\centering
		\includegraphics[clip,width=11cm]{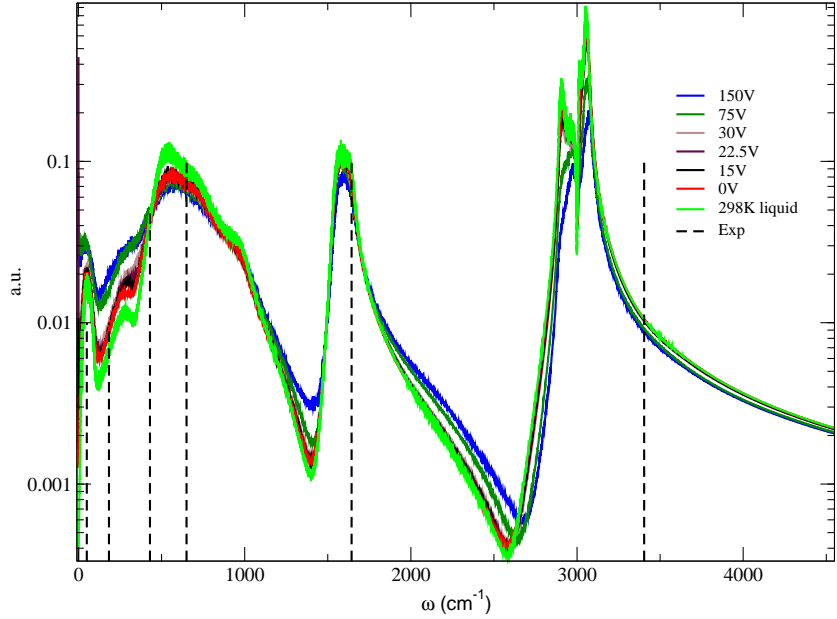}
		\caption{Infrared spectrum of FAB/$\varepsilon$ flexible model around different voltages is compared with the experimental value of bulk liquid at 298K and 1 bar of temperature and pressure respectively.}
		\label{Fig6}
	\end{figure}
	
	The changes in the intermolecular bonds length and angle can also be observed in the Infrared (IR) spectrum of the flexible model. Analyzing the behavior of the IR properties with respect to the increase of the electrical potential, we can see in figure~\ref{Fig6} that there are small differences that corresponds to the frequency at the maximum for the angle bending, the bond stretching and the bands. It's corresponds to the librational motion with symmetries A2, B2. As $V$ increases, the intermolecular stretch definition is lost as shown in Table~\ref{tableIR}. This leads to the molecule becoming less flexible and the cage vibrations, that are located at 50 cm$^{-1}$, becomes more defined as the applied tension increases.
	
	\begin{table}
		\caption{Experimental and simulation IR data results of FAB/$\varepsilon$ water model\cite{fbae}. In liquid phase at 298K and 1bar of temperature and pressure respectively. Wave numbers(cm $^{-1}$ ) at the peak of the bands of the power
			spectrum}
		\scalebox{0.8}[0.7]{
			\begin{tabular}{|l|c|c|}
				
				\hline		
				&Exp.&FBA/$\varepsilon$\\																		\hline
				\hline	
				Cage  vibrations	&	50		&	50		\\
				Intermolecular stretch	&	183.4	&	278		\\
				Librations  A2	&	430	&	536	\\
				Librations   B2	&	650	&	685	\\
				Bending  (H-O-H)	&	1643.5&	1613	\\
				Stretching  (O-H)	&	3404	&	3060	\\
				\hline		
		\end{tabular}}
		\label{tableIR}
	\end{table}
	
The polarization factor G$_k$ measures the equilibrium fluctuations of the collective dipole moment of the system and is related to the orientational correlation function. Kirkwood  postulate that it should be possible to express $\varepsilon$ in terms of a short-range orientational correlation function \cite{kirk}. In this way, here we evaluate the polarization factor G$_K$ \cite{Glattli},
\begin{equation}
\label{Ec5}
G_K= <{\bf M}^2> / N \mu^2
\end{equation}
 Here, $N$ is the total number of molecules, $M$ is the sum total of dipoles $ {\mu}$ in the system. Local orientational correlations are averaged out by thermal motion after the first few coordination shells.

\begin{figure}[t]
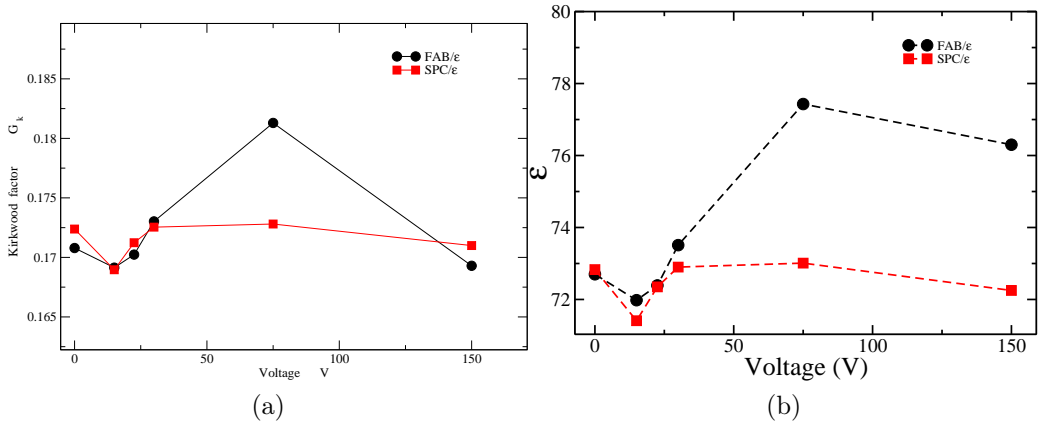

		\centering
	 \subfloat[]{\includegraphics[width=0.42\textwidth]{Gk.eps}}
		\subfloat[]{\includegraphics[width=0.41\textwidth]{eTotal-v.eps}}
	 \caption{(a) Kirkwood factor G$_k$ and (b) 
			Total dielectric constant as function of the electrical potential for the two H$_2$O models. Black circles stands for the FAB/$\varepsilon$ flexible model and red squares for the SPC/$\varepsilon$ rigid model.}
		\label{Fig6a}
	\end{figure}

Analyzing the polarization values G$_k$ and the average dielectric constant $\epsilon$ inside the pore as seen in the figure~\ref{Fig6a}, the polarization have a great increment up the V* and from there it decreases, as the dielectric constant.

	
	The changes in the water distinct dipole moment and the suppression of the molecule flexibility observed in the IR influences how the water  behaves as a dielectric medium. The static dielectric constant $\varepsilon$ of a polar liquid is related to the thermal equilibrium fluctuations of the polarization at zero field~\cite{fro}. Polarization fluctuations are long-range and vary with the shape of the dielectric body therefore, the confinement strongly affects the long-range interaction. We can see in the figure \ref{Fig6a} that both flexible and rigid water have a similar behavior. For low voltages the dielectric constant  decreases with voltage and reaches a minimum value. For voltages above $V_{min}$, the dielectric constant increases with voltage to a maximum, $V=V*$. For  $V>V*$ the dielectric constant decreases with  voltage. For the flexible model the maximum is located at $V^*$. The maximum values in $\varepsilon$ for the FAB/$\varepsilon$ model is higher than the observed for the rigid SPC/$\varepsilon$ water. This is a consequence from the fact that for rigid models, variations in the dielectric permitivity are related to distinct orientations of the molecules, changing the molecular arrangement. For flexible models, besides the orientations, the variation in the internal bonds length and angle leads to stronger voltage effects in the dielectric behavior. 
	
	Is interesting to compare the total dielectric constant in bulk and confined for both water models. For the rigid model SPC/$\varepsilon$, the bulk dielectric constant is $\varepsilon_{\rm bulk} = 78$~\cite{spce}. The comparison between this value and the dielectric contant of confined SPC/$\varepsilon$ as shown in  figure~\ref{Fig2} indicates that the confinement decreases the dielectric constant. This result agrees with previous computational and experimental studies~\cite{Lee84, Fumagalli18, Tocci14, Zhu21, Sato18, Varghese19} that indicates that water dipoles tends to orientate parallel to the confining surfaces, suppressing the parallel contribution to the total dielectric constant and the total dielectric constant itself. 
	
	For FAB/$\varepsilon$ water we observe a distinct scenario. This model have, in bulk, a dielectric constant of $\varepsilon = 76$. As we can see in the figure~\ref{Fig6a}, the total dielectric constant in the confined regime is smaller than the bulk value for low applied voltages. However, in the voltage regime where the dipole moment of the interfacial water molecules in regions I and III sharply increases, above 50V, the total dielectric constant increases, becoming higher than the bulk value. This shows that not only the orientation of the water molecule at the nanopore interface affects the dielectric constant, but at high voltages the internal bonds can play a major role.

\begin{figure}[t]
		\centering
		\includegraphics[clip,width=11cm]{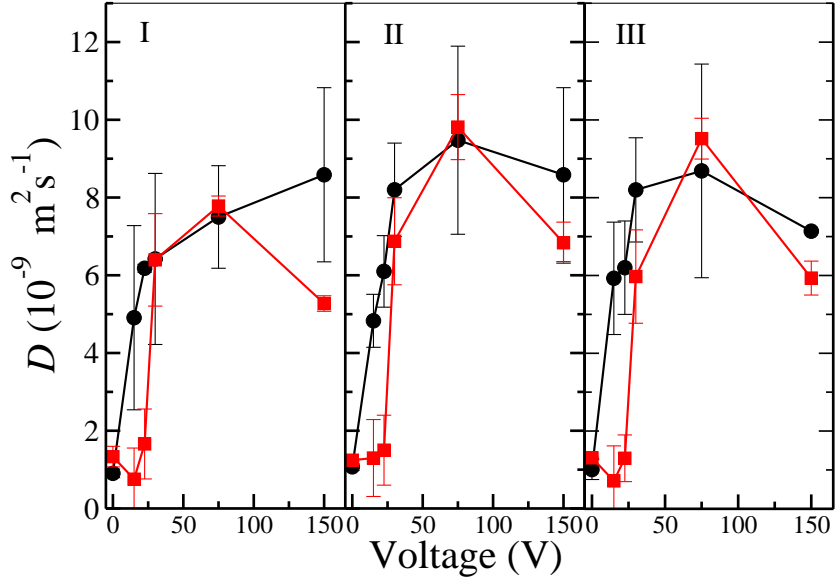}
		\caption{Diffusion coefficients for the H$_2$O confined between two
			charged walls calculated in the different regions defined in figure 1 and
			for different electric potentials. Black circles
			FAB/$\varepsilon$ flexible model and red squares with the SPC/$\varepsilon$ rigid model.}
		\label{Fig5}
	\end{figure}

	Finally, we analyze the dynamical behavior. Recent works have explored the relation between the dynamics of the contact layer and their electric properties. Experiments by Fumagalli and co-workers~\cite{Fumagalli18} have shown that the contact layer - with 3 molecules thickness - is electrically dead, with out-of-plane $\varepsilon$ of only 2. Using simulations of the rigid SPC/E water model, Mondal \textit{et al} observed that, despite electrically dead, this layers is not dynamically dead near a hydrophobic surface~\cite{Mondal20}. For our hydrophobic case, when the voltage is 0V, the interfacial layer for the rigid SPC/$\varepsilon$ water has a slightly higher diffusion than the flexible FBA/$\varepsilon$, as we can see in the figure~\ref{Fig5}, despite of have the same dielectric constant. Interesting, the diffusion of the contact layers for the SPC/$\varepsilon$ model reaches a minima at 22.5 V - like the dielectric constant. However, for the flexible model $D$ increases, diffusing faster than the non-confined case, whose diffusion constant is 2.39 $cm^2s^{-1}$ at 300 K and 1 bar. This can be related also to the dielectric constant: the FBA/$\varepsilon$ water $\varepsilon$ decreases less in comparison to the rigid model. As the voltage increases even more, both models achieves higher diffusion than the non-confined case  for the SPC/$\varepsilon$ model is 2.4 $cm^2s^{-1}$. This increase in the diffusion constant is followed by the increase in $\varepsilon$, reinforcing the relation between this two quantities for water under extreme confinement.

	\section{Conclusions}
	
	Molecular Dynamics simulations were carried out to study
	confinement of H$_2$O between two parallel graphene walls
	with a surface charge distribution, i.e a electric potential.
	Two different H$_2$O models were tested, the
	FAB/$\epsilon$ flexible and SPC/$\epsilon$ rigid models. 
	At zero electric potential both models present similar structure
	along the nano-capacitor and when the electric potential is increased
	a strong structure is formed close to the walls. In particular the
	FAB/$\epsilon$ model can adapt their structure and give us more information about this changes and if they are representative in the dielectric properties as the electric field increases.
	
	Due to the dipole moment induced at very strong electrical potentials, the dielectric constant changes within the nanopore and, therefore, the electrostatic properties are expected to change. We see a non-monotonic behavior of the total dielectric constant as function of the applied voltage, with a minima at 22.5 V. This minima is more pronounced for the rigid molecule, and is followed by a minima in the diffusion at the interfacial layer for this water model.The flexible molecule has a less pronounced minima in $\varepsilon$ and did not show a minima in $D$. This findings helps to elucidate the role of the water model to understand the behavior of fluids confined at extreme conditions.

\begin{acknowledgement}
	
	RFA acknowledge support from CONACYT, SECTIE, UAM  Supercomputing Center (Yotla) for the computer time allocated for some of our calculations.
	MCB and JRB thanks the Brazilian science agencies - Conselho
	Nacional de Desenvolvimento Cientıfico e Tecnologico (CNPq) (INCT-Fc) and Coordenac\=ao de Aperfeiçoamento de Pessoal
	de N\'ivel Superior (CAPES Print Program)

\end{acknowledgement}

\bibliography{achemso}

	\begin{figure}[]
		\centering
		\subfloat[]{\includegraphics[width=0.6\textwidth]{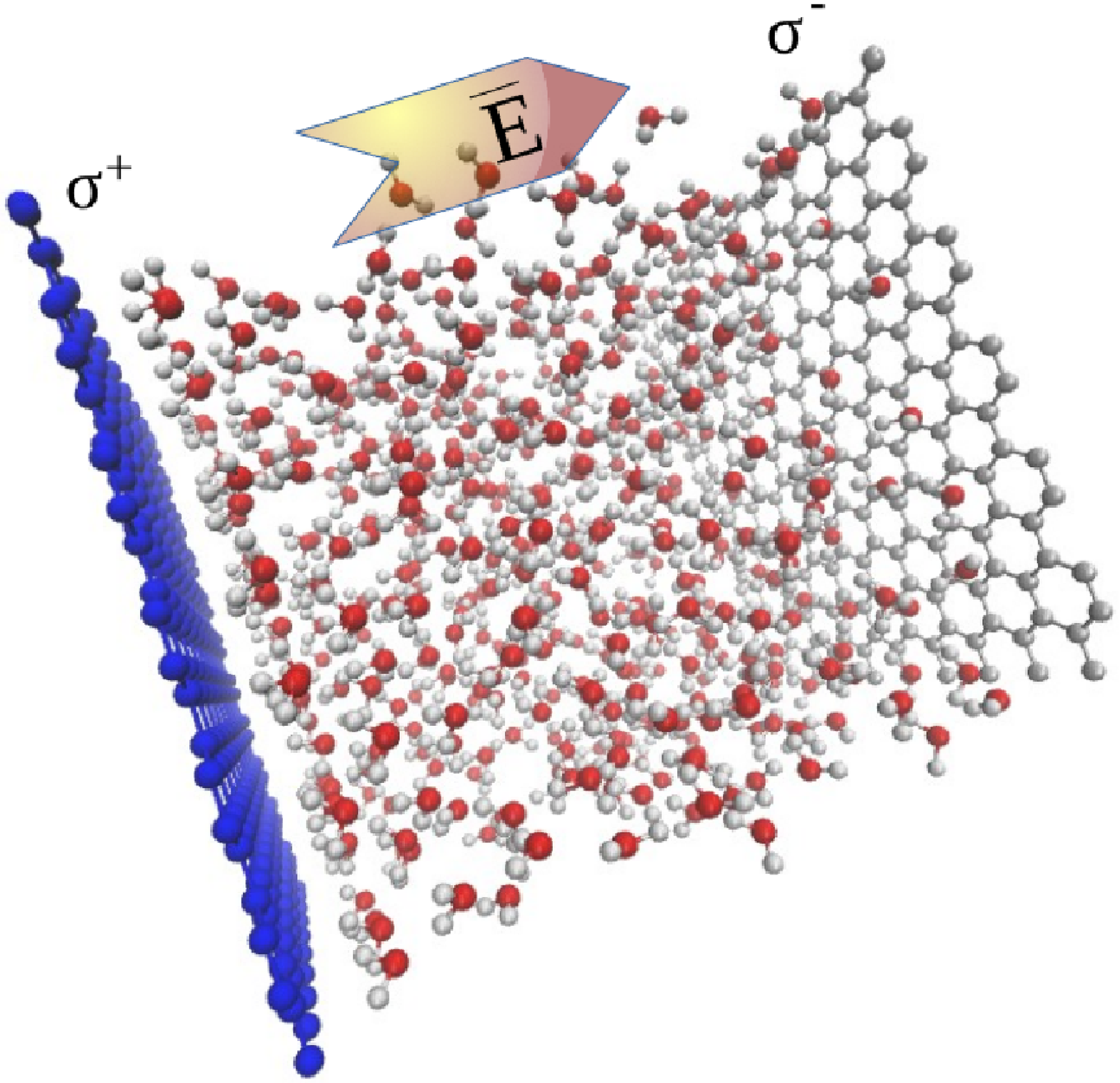}}
		\caption{TOC Figure}
		\label{FigA}
	\end{figure}

\end{document}